\documentclass[aps,prd,preprint,superscriptaddress]{revtex4-1}
\usepackage[mathscr]{euscript}
\usepackage{float,epsfig}
\usepackage{bm}% bold math
\usepackage{graphicx}% Include figure files
\usepackage{subfigure}
\usepackage{amsmath,amssymb,amsthm}
\usepackage[colorlinks=true,linkcolor=blue,urlcolor=blue]{hyperref}
\newcommand{\bea}{\begin{eqnarray}}
\newcommand{\eea}{\end{eqnarray}}
\newcommand{\beq}{\begin{equation}}
\newcommand{\eeq}{\end{equation}}

\def\/{\over}

\begin{document}

%\preprint{}

\title{Nonadditive quantum gravitational interaction for three nonpointlike objects}


\author{Yongshun Hu}
\email[]{huyongshun@ucas.ac.cn}
\affiliation{School of Fundamental Physics and Mathematical Sciences, Hangzhou Institute for Advanced Study, UCAS, Hangzhou 310024, China}
\affiliation{School of Physical Sciences, University of Chinese Academy of Sciences, Beijing 100049, China}
%\author{Jiawei Hu}
%\email[Corresponding author. ]{jwhu@hunnu.edu.cn}
%\affiliation{Department of Physics and Synergetic Innovation Center for Quantum Effects and Applications, Hunan Normal University, Changsha, Hunan 410081, China}
\author{Hongwei Yu}
\email[Corresponding author. ]{hwyu@hunnu.edu.cn}
\affiliation{Department of Physics and Synergetic Innovation Center for Quantum Effects and Applications, Hunan Normal University, Changsha, Hunan 410081, China}

%\thanks{}


%\date{\today}

\begin{abstract}

We explore the nonadditive three-body quantum gravitational quadrupole interaction among three nonpointlike objects in their ground states in the framework of linearized quantum gravity and find that the interaction exhibits a distance behavior of $r_A^{-5}r_B^{-5}r_C^{-5}$ in the near regime where the interobject distances are small compared with the characteristic transition wavelength of the objects, and  $r_A^{-5}r_B^{-5}r_C^{-5}(r_A+r_B+r_C)^{-1}$ in the far regime where the distances are larger than the characteristic transition wavelength, where $r_\xi $ denotes the distance between any two objects except object $\xi\;(\xi=A,B,C)$.
Compared to the additive quantum gravitational interaction between two ground-state objects in vacuum which is always attractive both in the near and far regimes, the nonadditive interaction among three ground-state objects can be either attractive or repulsive depending on the geometrical configuration of these objects. That is, in principle, the attractive or repulsive properties of the nonadditive three-body quantum gravitational interaction can be manipulated by changing the geometrical arrangement of the objects.

\end{abstract}

%\keywords{}

\maketitle

\section{Introduction}
\label{sec_in}
\setcounter{equation}{0}
%%%%%%%%%%%%%%%%%%%%%%%%%%%%%%%%%%%%%%%%%%%%%%%%%%
In a quantum sense, there inevitably exist quantum fluctuations of fields in vacuum, which may induce observable effects. In general, fluctuating fields in vacuum induce instantaneous multipole moments in atoms or objects, which then interact  via the exchange of virtual mediating particles, and an interaction energy is thus resulted. One of the well-know examples is the electromagnetic Casimir-Polder (CP) interaction, which arises from the vacuum-fluctuation-induced dipole-dipole interaction between two neutral atoms via the exchange of virtual photons \cite{CP}.
%Physically speaking, electromagnetic vacuum fluctuations induce instantaneous electric dipole moments in two neutral atoms, which then couple with each other via the exchange of virtual photons.
Such interatomic CP effects have been widely investigated in various circumstances, e.g., in the presence of external electromagnetic fields \cite{Thirunamachandran1980,Milonni1992,Milonni1996,Bradshaw2005,Andrews2006,Salam2006, Salam2007,Sherkunov2009,Mackrodt1974,Passante2020,yongs2021pra}, thermal fluctuations \cite{Goedecke1999,Ninham1999}, and  boundaries \cite{Passante2006,power1982}.
Also, it has been found that the interatomic CP interactions behave differently in terms of distance-dependence when the two atoms are in different quantum states~\cite{PT,Salam,Power1993pra,McLone1965,Gomberoff1966, Power1993Chem,Power1995,Rizzuto2004, Sherkunov2007, Preto2013,Donaire2015,Milonni2015,Berman2015,Jentschura2017}.
%Also, such interactions can be significantly affected by an external electromagnetic wave as well as an electrostatic field \cite{Thirunamachandran1980,Milonni1992,Milonni1996,Bradshaw2005,Andrews2006,Salam2006, Salam2007,Sherkunov2009,Mackrodt1974,Passante2020,yongs2021pra}.

Likewise, in the gravitational case, there should also exist vacuum-fluctuation-induced quantum gravitational quadrupole-quadrupole interactions between nonpointlike objects, if one accepts that basic quantum principles are  applicable to gravity as well. However, a full theory of quantum gravity has not yet been established. Nonetheless, one can still investigate low energy quantum gravitational effects in the framework of the effective field theory or linearized quantum gravity.  Such investigations are based on the belief that the results obtained should agree with those  in a full theory of quantum gravity at low energy scales.  %Recently, such CP-like quantum gravitational quadrupole-quadrupole interactions have been studied in Refs. \cite{Ford2016,Wu2016,Wu2017,Holstein2017,yu2018,yongs2020epjc,yongs2020prd,yongs2021prd}.
Similar to the electromagnetic case, the gravitational CP-like quadrupole-quadrupole interactions are found to be dependent on the quantum states of the two-object systems \cite{Ford2016,Wu2016,Wu2017,Holstein2017,yongs2020epjc} and can be modified by the thermal fluctuations of gravitons \cite{Wu2017}, the gravitational boundaries \cite{yu2018}, and the external gravitational radiation fields \cite{yongs2020prd,yongs2021prd}.
For example, the vacuum-fluctuation-induced gravitational quadrupole-quadrupole potential for two ground-state nonpointlike objects behaves as $r^{-10}$ and $r^{-11}$ in the near and far regimes respectively, while for two entangled objects it behaves as $r^{-5}$ and $r^{-1}$ in the near and far regimes respectively.
The presence of gravitational boundaries or a thermal bath of gravitons causes temperature-dependent or boundary-dependent modifications to the interaction potentials in vacuum \cite{Wu2017,yu2018}.
Moreover, in the presence of external gravitational radiation fields, the induced interobject quantum gravitational quadrupole-quadrupole interaction can be  attractive or repulsive depending on the properties (e.g., polarization et al.) of the external gravitational fields \cite{yongs2020prd,yongs2021prd}, which is significantly different from that of the vacuum case which is always attractive.

The CP-like quantum gravitational interactions discussed above are obviously additive, since %only two nonpointlike objects are taken into account and
the corresponding interaction potentials are related just to the distance between the pair. Naturally, a question arises as to whether the CP-like quantum gravitational interaction is still additive in the presence of a third nonpointlike object, or in other words, whether a nonadditive three-body quantum gravitational quadrupole interaction exists.
Again, there are similar examples in quantum electrodynamics.
It has been demonstrated that the electromagnetic CP interaction between two neutral atoms can be modified by the presence of a third one which then generates a triple-dipole interaction potential and gives rise to a nonadditive three-atom interaction~\cite{A-T1943,Aub1960,McLachlan1963,PT1985,Cirone1997,Passante1998}. %\cite{A-T1943,Aub1960,McLachlan1963,PT1985,PT1994,Cirone1997,Passante1998}.
Physically speaking, the electromagnetic three-body interaction among three neutral atoms in vacuum arises from the processes of three-photon exchange wherein only one photon is exchanged between each pair of atoms, so that the interaction potential is related to distances between every pair of atoms and is thus nonadditive. Such correlated interaction processes among three atoms are clearly different from those of the two-atom case, which arise from two-photon exchange between the pair and from which the resulting interaction is obviously additive.
Similarly, in the gravitational case, one may also expect a nonadditive three-body quantum gravitational quadrupole interaction when a third nonpointlike object is present.

In this paper, we explore the nonadditive three-body quantum gravitational quadrupole interaction among three nonpointlike objects in their ground states based on the theory of linearized quantum gravity. First, we derive the three-body quantum gravitational potential in the framework of linearized quantum gravity. Then, we discuss its behaviors in specific cases and show concretely a geometrical dependence of the attractive or repulsive properties of the interaction. Throughout this paper, the Latin indices and the Greek indices run from $1$ to $3$ and $0$ to $3$, respectively. The Einstein summation convention for repeated indices is assumed and units with $\hbar=c=16\pi G=1$ are used, where $\hbar$ is the reduced Planck constant, $c$ is the speed of light and $G$ is the Newtonian gravitational constant.

%%%%%%%%%%%%%%%%%%%%%%%%%%%%%%%%%%%%%%%%%%%%%%%%%%
\section{Basic equations}
\label{sec_ba}
%\setcounter{equation}{0}
%%%%%%%%%%%%%%%%%%%%%%%%%%%%%%%%%%%%%%%%%%%%%%%%%%
We consider three nonpointlike objects (labeled as A, B and C) coupled with the fluctuating gravitational fields in vacuum. For simplicity, let us model these objects by two-level systems with the ground and excited states being $|g_\xi\rangle$ and $|e_\xi\rangle$ respectively, and label the corresponding energy spacing as $\omega_\xi$ ($\xi=A,B,C$). The total Hamiltonian of the system can be written as
\beq \label{Hamiltonian}
H=H_F+H_S+H_I\;,
\eeq
where $H_{F}$ is the Hamiltonian of the gravitational fields, $H_{S}$ represents the Hamiltonian of the three nonpointlike objects (A, B and C), and $H_I$ denotes the interaction Hamiltonian between the nonpointlike objects and the gravitational fields, which takes the form
\beq\label{HI}
H_I=-\frac{1}{2}Q^{A}_{ij}E_{ij}(\vec x_A)-\frac{1}{2}Q^{B}_{ij}E_{ij}(\vec x_{B})-\frac{1}{2}Q^{C}_{ij}E_{ij}(\vec x_{C})\;,
\eeq
where $Q^{\xi}_{ij}$ is the quadrupole moment operator of object $\xi$ ($\xi=A,B,C$),  and $E_{ij}$ is the gravitoelectric tensor of the fluctuating gravitational fields in vacuum, which is defined as $E_{ij}=C_{0i0j}$ by an analogy between the linearized Einstein field equations and the Maxwell equations~~\cite{Campbell1971,Campbell1976,Maartens1998,Matte1953,Ramos2010,Szekeres1971,Ruggiero2002}, where $C_{\mu\nu\alpha\beta}$ is the Weyl tensor. Under the weak-field approximation, the spacetime metric can be written as a sum of the flat spacetime metric and a linearized perturbation $h_{\mu\nu}$. Then, the gravitoelectric tensor $E_{ij}$ is found to be
\beq
E_{ij}=\frac{1}{2}\ddot h_{ij}\;.
\eeq
In the transverse traceless gauge, the linearized metric perturbation can be quantized as~\cite{yu1999}
\beq
h_{ij}=\sum_{\vec p, \lambda}\sqrt{\frac{1}{2\omega(2\pi)^3}}[ a_{\lambda}(\omega)e^{(\lambda)}_{ij} e^{i(\vec{p}\cdot\vec{x}-\omega t)}+\text{H.c.} ]\;,
\eeq
where $a_{\lambda}(\omega)$ is the annihilation operator of the fluctuating gravitational fields, $e^{(\lambda)}_{ij}$ are polarization tensors, $\lambda$ labels the polarization states, $\omega=|\vec p|= (p_{x}^2+p_{y}^2+p_{z}^2)^{1/2}$, and H.c. denotes the Hermitian conjugate.

\begin{figure}[htbp]
  \centering
   %Requires \usepackage{graphicx}
  \includegraphics[width=0.5\textwidth]{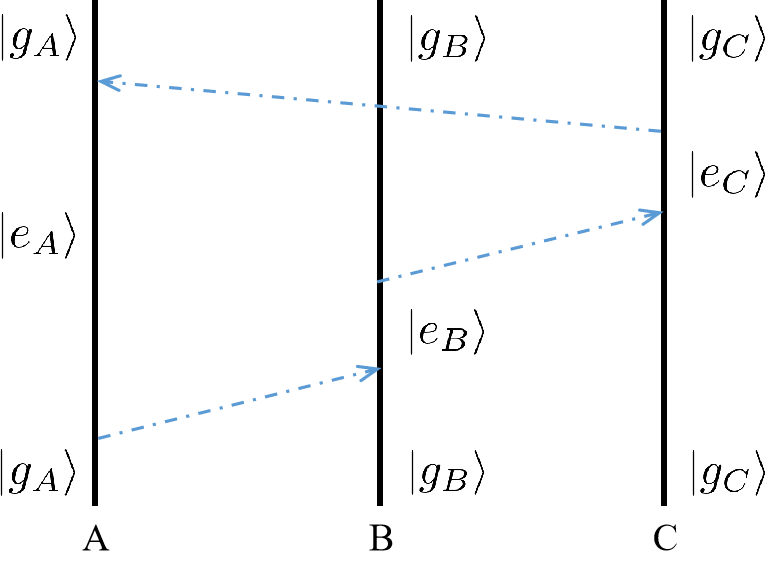}\\
  \caption{One typical time-ordered diagram for the calculation of the nonadditive three-body quantum gravitational interaction. The blue dotted line represents a virtual graviton. }\label{TT1}
\end{figure}

\begin{figure}[htbp]
  \centering
   %Requires \usepackage{graphicx}
  \includegraphics[width=0.75\textwidth]{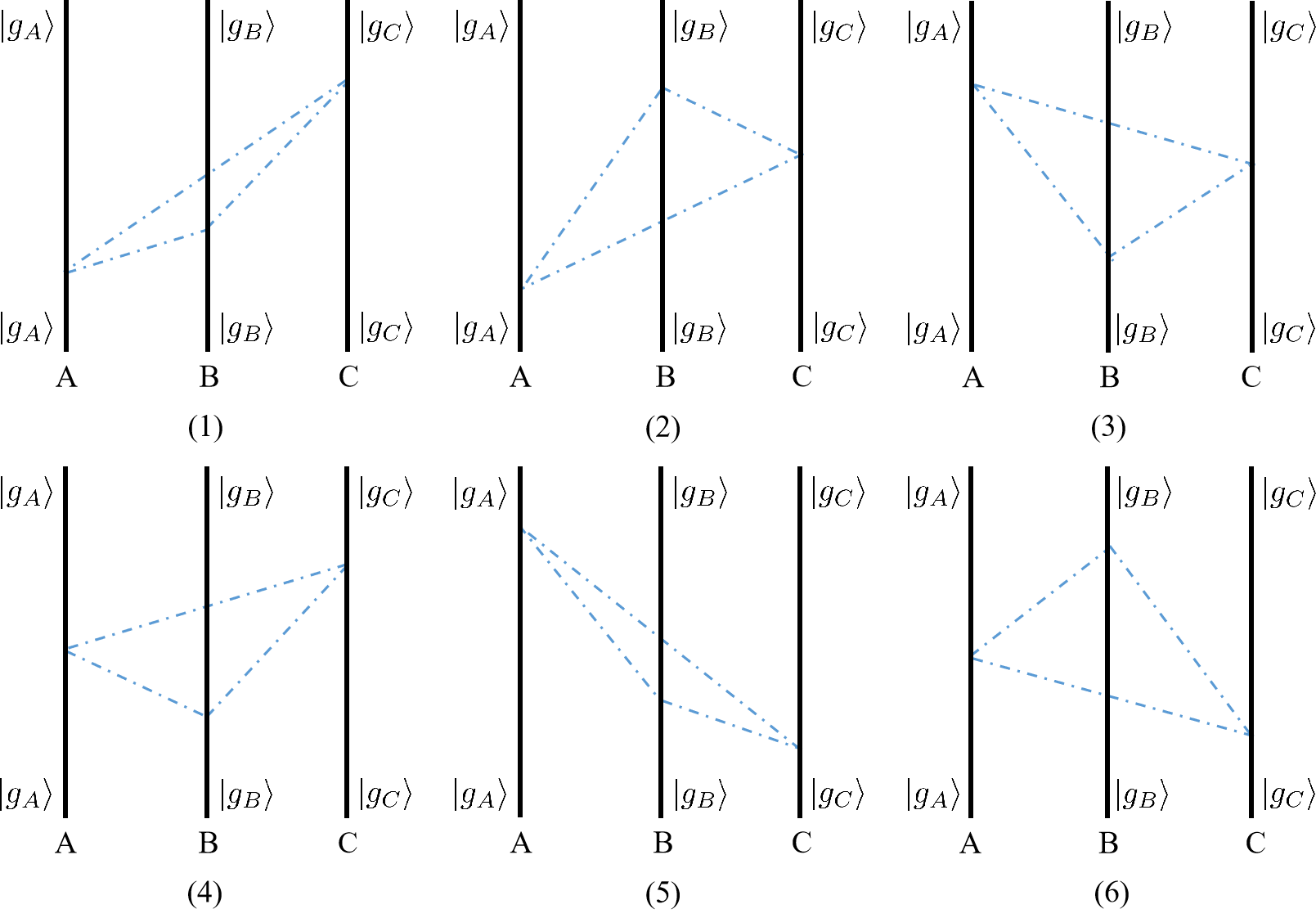}\\
  \caption{Time-ordered diagrams contributing to three-body quantum gravitational interaction when an effective two-graviton coupling Hamiltonian is employed. }\label{F6}
\end{figure}

We choose the initial state of the whole system as
\beq\label{phi}
|\phi\rangle=|g_A\rangle |g_B\rangle |g_C\rangle|0\rangle\;,
\eeq
and denote its energy as $E_{\phi}$, where $|0\rangle$ represents the vacuum state of the fluctuating gravitational field. In analogy to the electromagnetic case \cite{Aub1960}, the nonadditive three-body quantum gravitational interaction energy  can be obtained from  the sixth-order perturbation theory, which contains $360$ possible time-ordered diagrams, and a typical one is shown in Fig.~\ref{TT1}.
However, a direct sixth-order calculation is exceedingly unwieldy and, for convenience, simplifications of the interaction processes are needed. For this purpose, an effective two-graviton interaction Hamiltonian is introduced, and for the first step, we re-express the vacuum-fluctuation-induced quadrupole in object $\xi$ as
\beq\label{Qij}
Q^{\xi}_{ij}=\frac{1}{2}\alpha^{\xi}_{ijkl}(\omega ) E_{kl}\;,
\eeq
where $\alpha^{\xi}_{ijkl}(\omega )$ is the gravitational polarizability of object $\xi$, which takes the form
\beq
\alpha^{\xi}_{ijkl}(\omega )=\frac{2\omega_\xi \hat Q^{\xi}_{ij}\hat Q^{\xi}_{kl}}{\omega_\xi^2-\omega^2}\;,
\eeq
where $\hat Q^{\xi}_{ij}$ represents the quadrupole transition moments. Then, the effective Hamiltonian can be written as
\beq\label{Heff}
H^{eff}_I=-\frac{1}{4}\sum_{\xi=A,B,C} \alpha^{\xi}_{ijkl}(\omega ) E_{ij}(\vec x_\xi) E_{kl}(\vec x_\xi)\;.
\eeq
The three-body quantum gravitational interaction energy shift can  be calculated based on the third-order perturbation theory
\beq
\Delta E=\sum_{I,II}\frac{\langle \phi|H^{eff}_{I}|II\rangle\langle II|H^{eff}_{I}|I\rangle\langle I|H^{eff}_{I}|\phi\rangle}{(E_{II}-E_{\phi})(E_{I}-E_{\phi})}\;,
\eeq
with only six contributing time-ordered diagrams as shown  in Fig. \ref{F6}.
%For these four diagrams, the  intermediate state $|I\rangle=|g_A g_B\rangle|1_{\vec p \lambda}\rangle|N-1\rangle$ correspond to $(a)(c)$, while $|I\rangle=|g_A g_B\rangle|1_{\vec p \lambda}\rangle|N+1\rangle$ correspond to $(b)(d)$.
Summing up all the contributions, the three-body gravitational interaction is found to be
\bea\label{Eabc 1}
\nonumber \Delta E_{ABC}&=&-\frac{1}{2^{15}(2\pi)^9} \sum_{\vec p_1,\lambda_1} \sum_{\vec p_2,\lambda_2}
\sum_{\vec p_3,\lambda_3} [\alpha^{A}_{ijkl}(\omega_1)+\alpha^{A}_{ijkl}(\omega_3)] [\alpha^{B}_{ghmn}(\omega_1)+\alpha^{B}_{ghmn}(\omega_2)]  \\
\nonumber&&\times [\alpha^{C}_{efqs}(\omega_2)+\alpha^{C}_{efqs}(\omega_3)]e^{(\lambda_1)}_{ij}e^{(\lambda_3)}_{kl} e^{(\lambda_1)}_{gh}e^{(\lambda_2)}_{mn} e^{(\lambda_2)}_{ef}e^{(\lambda_3)}_{qs} e^{i \vec p_1 \cdot \vec r_C} e^{i \vec p_2 \cdot \vec r_A} e^{i \vec p_3 \cdot \vec r_B} \\
&&\times \left[\frac{2}{(\omega_1+\omega_3)(\omega_1+\omega_2)}+ \frac{2}{(\omega_1+\omega_2)(\omega_2+\omega_3)}+\frac{2}{(\omega_1+\omega_3)(\omega_2+\omega_3)}\right]\;,
\eea
where $\vec r_A= \vec x_C - \vec x_B$, $\vec r_B= \vec x_A - \vec x_C$, and $\vec r_C= \vec x_B - \vec x_A$ have been defined. Using the summation of polarization tensors in the transverse traceless gauge~\cite{yu1999}
\bea
\nonumber\sum_{\lambda}e^{(\lambda)}_{ij}e^{(\lambda)}_{kl}&=&\delta_{ik}\delta_{jl}+\delta_{il}\delta_{jk} -\delta_{ij}\delta_{k l} +\hat p_i \hat p_j \delta_{kl}+\hat p_k \hat p_l \delta_{ij}\\
&&-\hat p_i\hat p_k\delta_{jl}- \hat p_i \hat p_l \delta_{jk}-\hat p_j \hat p_k \delta_{il} - \hat p_j \hat p_l \delta_{ik}+ \hat p_i \hat p_j \hat p_k\hat p_l \;,
\eea
where $ \hat p_i$ is the $i$-th component of the unit vector $\vec p/p$,  we obtain
\bea\label{H_ijkl}
\nonumber \sum_{\lambda}e^{(\lambda)}_{ij}e^{(\lambda)}_{kl} e^{i \vec p \cdot \vec r} &=&\frac{1}{\omega^4}\Big[(\delta_{ik}\delta_{jl}+\delta_{il}\delta_{jk} -\delta_{ij}\delta_{k l})\nabla^4 +(\partial_i \partial_j \delta_{kl}+\partial_k \partial_l \delta_{ij}\\
&&\nonumber -\partial_i\partial_k\delta_{jl}- \partial_i \partial_l \delta_{jk}-\partial_j \partial_k \delta_{il} - \partial_j \partial_l \delta_{ik})\nabla^2+ \partial_i \partial_j \partial_k \partial_l \Big]e^{i \vec p \cdot \vec r}\\
&=&\frac{1}{\omega^4}\hat H_{ijkl}^{r} e^{i \vec p \cdot \vec r}\;,
\eea
and
\bea
\int \frac{1}{\omega^4}\hat H_{ijkl}^{r} e^{i \vec p \cdot \vec r} d\Omega_{\vec p} =4\pi \hat H_{ijkl}^{r} \frac{\sin{\omega r}}{\omega^5 r}\;,
\eea
where $\hat H_{ijkl}^{r}$ is a differential operator whose definition follows directly from Eq. (\ref{H_ijkl}), $\nabla^2=\partial_i\partial^i$, and $\Omega_{\vec p}$ denotes the solid angle. Replacing the summation over wave vectors and polarization states in Eq. (\ref{Eabc 1}) by integration and performing the integral with the help of the integral representation
\bea
\nonumber&&\frac{1}{(\omega_1+\omega_3)(\omega_1+\omega_2)}+\frac{1}{(\omega_1+\omega_2)(\omega_2+\omega_3)}+ \frac{1}{(\omega_1+\omega_3)(\omega_2+\omega_3)}\\
&&=\frac{4}{\pi}\int^{\infty}_{0}\frac{\omega_1\omega_2\omega_3 d u} {(\omega_1^2+u^2)(\omega_2^2+u^2)(\omega_3^2+u^2)}\;,
\eea
we obtain
\bea\label{Eabc 2}
\nonumber \Delta E_{ABC}&=&-\frac{1}{2^{9}(2\pi)^6}\hat H_{mnef}^{r_A}\hat H_{klqs}^{r_B}\hat H_{ijgh}^{r_C} \int_{0}^{\infty}d \omega_1\int_{0}^{\infty}d \omega_2 \int_{0}^{\infty} d \omega_3 \Big[\alpha^{A}_{ijkl}(\omega_1)+\alpha^{A}_{ijkl}(\omega_3)\Big]   \\
\nonumber&&\times \Big[\alpha^{B}_{ghmn}(\omega_1)+\alpha^{B}_{ghmn}(\omega_2)\Big]\Big[\alpha^{C}_{efqs}(\omega_2) +\alpha^{C}_{efqs}(\omega_3)\Big] \frac{\sin{\omega_2 r_A}\sin{\omega_3 r_B}\sin{\omega_1 r_C}}{\pi r_A r_B r_C}\\
&&\times \int^{\infty}_{0}\frac{\omega_1\omega_2\omega_3 d u} {(\omega_1^2+u^2)(\omega_2^2+u^2)(\omega_3^2+u^2)}\;.
\eea
To evaluate the frequency integrals in Eq. (\ref{Eabc 2}), an integral representation is utilized, i.e.,
\bea
\int_{0}^{\infty}\alpha^\xi_{ijkl}(p)\frac{p \sin{p r}}{p^2+u^2} dp=\frac{1}{4 i}\int_{-\infty}^{\infty} \alpha^\xi_{ijkl}(p) e^{i p r}\left(\frac{1}{p+i u}+\frac{1}{p-i u} \right)dp= \frac{\pi}{2}\alpha^\xi_{ijkl}(i u)e^{-u r}\;.
\eea
%in analogy to the electromagnetic case~\cite{Passante1998}.
Then, the expression of the nonadditive three-body quantum gravitational quadrupole interaction is finally obtained
\bea\label{Eabc final}
\Delta E_{ABC}&=&-\frac{1}{2^{15}\pi^4}\hat H_{mnef}^{r_A}\hat H_{klqs}^{r_B}\hat H_{ijgh}^{r_C} \int_{0}^{\infty} \alpha^{A}_{ijkl}(i u) \alpha^{B}_{ghmn}(i u) \alpha^{C}_{efqs}(i u) \frac{e^{-u (r_A+r_B+r_C)}}{r_A r_B r_C}d u\;.
\eea

\begin{figure}[htbp]
  \centering
   %Requires \usepackage{graphicx}
  \includegraphics[width=0.5\textwidth]{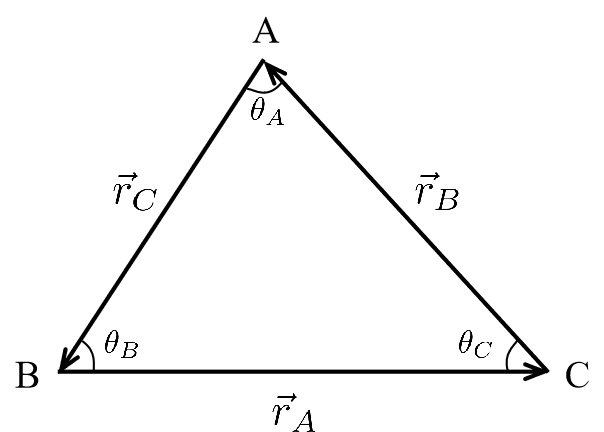}\\
  \caption{The notations for the vertices and interobject distance vector used for three objects.}\label{Triangle}
\end{figure}

Now  discussions about the three-body interaction potential Eq. (\ref{Eabc final}) are in order. For simplicity,  we assume that the objects A, B, and C are isotropically polarizable so that
\beq
\alpha^{\xi}_{ijkl}(\omega)=(\delta_{ik}\delta_{jl}+\delta_{il}\delta_{jk})\alpha^{\xi}(\omega)\;,
\eeq
where $\alpha^{\xi}(\omega)$ denotes the isotropic polarizability of object $\xi$. Then, Eq. (\ref{Eabc final}) can be simplified as
\bea\label{Eabc iso}
\Delta E_{ABC}&=&-\frac{1}{2^{12}\pi^4}\hat H_{klmn}^{r_A}\hat H_{ijmn}^{r_B}\hat H_{ijkl}^{r_C} \int_{0}^{\infty} \alpha^{A}(i u) \alpha^{B}(i u) \alpha^{C}(i u) \frac{e^{-u (r_A+r_B+r_C)}}{r_A r_B r_C}d u\;.
\eea
First, in the near regime, where the interobject distance $r_A$, $r_B$, and $r_C$ are all small compared with the characteristic transition wavelength of the objects, the integral in Eq. (\ref{Eabc iso}) is effectively limited to the region where $e^{-u(r_A+r_B+r_C)}\approx1$ and the asymptotic result is then obtained as
\bea
\Delta E_{ABC}&\simeq&-\frac{1}{2^{12}\pi^4}\hat H_{klmn}^{r_A}\hat H_{ijmn}^{r_B}\hat H_{ijkl}^{r_C} \frac{1}{r_A r_B r_C} \int_{0}^{\infty} \alpha^{A}(i u) \alpha^{B}(i u) \alpha^{C}(i u)d u\;.
\eea
Notice that
\bea
\nonumber\hat H_{klmn}^{r}\frac{1}{r}&=& \frac{1}{r^5} [3(\delta_{kl}\delta_{mn}+\delta_{km}\delta_{ln} +\delta_{kn}\delta_{lm}) -15(\hat r_k \hat r_l\delta_{mn}+\hat r_m \hat r_n\delta_{kl}\\
&&+\hat r_k \hat r_m\delta_{ln}+\hat r_k \hat r_n\delta_{lm}+\hat r_l \hat r_m\delta_{kn}+\hat r_l \hat r_n \delta_{km})+105\hat r_k \hat r_l \hat r_m \hat r_n]\;,
\eea
where $\hat r_n$ is the $n$-th component of the unit vector $\vec r/r$ ($r=r_A$, $r_B$, $r_C$).
After some algebraic manipulations, we obtain the near-zone three-body quantum gravitational interaction potential
\bea\label{Eabc near}
\Delta E_{ABC}&\simeq&-\frac{405}{2^{12}\pi^4 r_A^5 r_B^5 r_C^5} F(\theta_A,\theta_B,\theta_C) \int_{0}^{\infty} \alpha^{A}(i u) \alpha^{B}(i u) \alpha^{C}(i u)d u\;,
\eea
with
\bea
\nonumber F(\theta_A,\theta_B,\theta_C)&=&18 +110\cos{\theta_A}\cos{\theta_B}\cos{\theta_C} - 35(\cos^2{\theta_A}\cos^2{\theta_C} + \cos^2{\theta_B}\cos^2{\theta_C} \\
&& +\cos^2{\theta_A}\cos^2{\theta_B})+ 245\cos^2{\theta_A}\cos^2{\theta_B}\cos^2{\theta_C}\;,
\eea
where $\theta_A$, $\theta_B$, and $\theta_C$ are the internal angles of the triangle which are opposite to sides $BC$, $CA$, and $AB$, respectively. A schematic diagram of the geometrical configuration for the three objects is shown in Fig. \ref{Triangle}.
Here $\cos{\theta_A}=-\hat r_{B}\cdot \hat r_{C}$, $\cos{\theta_B}=-\hat r_{A}\cdot \hat r_{C} $ and $\cos{\theta_C}=-\hat r_{A}\cdot \hat r_{B} $ are used, and the identity for internal angles of a triangle, i.e.,
\beq
\cos^2{\theta_A}+\cos^2{\theta_B}+\cos^2{\theta_C}=1-2\cos{\theta_A}\cos{\theta_B}\cos{\theta_C}\;,
\eeq
has been applied. Notice that, the function $F(\theta_A,\theta_B,\theta_C)$ can be either positive or negative depending on the geometrical configuration,  % $\theta_A$, $\theta_B$, and $\theta_C$,
and as a result the interaction can be attractive or repulsive.  %That is, in the near regime, the three-body quantum gravitational quadrupole interaction depends on the geometrical configuration of the three objects.
Several concrete examples are helpful to illustrate such a geometrical dependence or influence. For a linear arrangement of the three objects with $r_A=2r_B=2r_C=2r$, $\theta_A=\pi$, and $\theta_B=\theta_C=0$, the three-body interaction potential Eq. (\ref{Eabc near}) is
\beq\label{Eabc near1}
\Delta E_{ABC}\simeq-\frac{1215}{2^{13}\pi^4 r^{15}}\int_{0}^{\infty} \alpha^{A}(i u) \alpha^{B}(i u) \alpha^{C}(i u)d u\;.
\eeq
So, the three-body quantum gravitational quadrupole interaction exhibits an $r^{-15}$ dependence in the near regime. Since the sign of the potential Eq. (\ref{Eabc near1}) is negative, the interaction is attractive for the case of a linear geometrical arrangement. For non-linear arrangements of the three objects, we now consider special two cases, which give rise to a positive and a negative value of $F(\theta_A,\theta_B,\theta_C)$ respectively.
For an equilateral triangle in which $r_A=r_B=r_C=r$ with $\theta_A=\theta_B=\theta_C=\frac{\pi}{3}$, the interaction potential Eq. (\ref{Eabc near}) is
\beq\label{Eabc near2}
\Delta E_{ABC}\simeq-\frac{752085}{2^{18}\pi^4 r^{15}}\int_{0}^{\infty} \alpha^{A}(i u) \alpha^{B}(i u) \alpha^{C}(i u)d u\;,
\eeq
while for a triangle in which $r_A=\sqrt{3}r_B=\sqrt{3}r_C=\sqrt{3}r$ with $\theta_A=\frac{2}{3}\pi$, $\theta_B=\theta_C=\frac{\pi}{6}$, the potential Eq. (\ref{Eabc near}) is
\beq\label{Eabc near3}
\Delta E_{ABC}\simeq\frac{20745\sqrt{3}}{2^{18}\pi^4 r^{15}}\int_{0}^{\infty} \alpha^{A}(i u) \alpha^{B}(i u) \alpha^{C}(i u)d u\;.
\eeq
Obviously, the distance dependence of such a three-body interaction in the triangle configurations is also $r^{-15}$. Notice that the sign of Eq. (\ref{Eabc near2}) is negative while that of Eq. (\ref{Eabc near3}) is positive, which respectively implies an attractive and a repulsive force associated with the corresponding geometrical arrangement of the three objects. That is, in the near regime, the attractive or repulsive properties of the three-body quantum gravitational potential depend crucially on the geometrical configuration of the three objects.

Second, in the far regime, where the interobject distance $r_A$, $r_B$, and $r_C$ are all larger than the characteristic transition wavelength of the objects so that the polarizability $\alpha^{\xi}(i u)$ can be approximately replaced by the static one (i.e., $\alpha^{\xi}(0)$), since, in the present case, small values of $u$ could provide the dominating contribution due to the existence of an exponential in the integrand in Eq. (\ref{Eabc iso}), we obtain
\bea
\nonumber \Delta E_{ABC}&\simeq&-\frac{1}{2^{12}\pi^4}\alpha^{A}(0) \alpha^{B}(0) \alpha^{C}(0)\hat H_{klmn}^{r_A}\hat H_{ijmn}^{r_B}\hat H_{ijkl}^{r_C}\int_{0}^{\infty} \frac{e^{-u(r_A+r_B+r_C)}}{r_A r_B r_C} du \\
&=&-\frac{1}{2^{12}\pi^4}\alpha^{A}(0) \alpha^{B}(0) \alpha^{C}(0)\hat H_{klmn}^{r_A}\hat H_{ijmn}^{r_B}\hat H_{ijkl}^{r_C}\frac{1}{r_A r_B r_C (r_A+r_B+r_C)}\;. \label{Eabc far}
\eea
Note that a detailed calculation of the derivative in Eq. (\ref{Eabc far}) is complicated and tedious, and it is also profitless to take up much space to show such an expression. For simplicity, here we only give several special examples with certain geometrical configurations of the three objects, which may be enough to show the essential properties of the three-body quantum gravitational potential in the far regime. Also, a serviceable derivative may be helpful, which is given as follows
\bea
\nonumber\hat H_{klmn}^{r}\frac{e^{-u r}}{r}&=& \frac{e^{-u r}}{r^5} [U_1(u r)(\delta_{km}\delta_{ln}+ \delta_{kn}\delta_{lm}) + U_2(u r)\delta_{kl}\delta_{mn}+ U_3(u r)(\hat r_k \hat r_l\delta_{mn}+\hat r_m \hat r_n \delta_{kl})\\
&&+U_4(u r)(\hat r_k \hat r_m\delta_{ln}+\hat r_k \hat r_n\delta_{lm}+\hat r_l \hat r_m\delta_{kn}+\hat r_l \hat r_n \delta_{km})+ U_5(u r)\hat r_k \hat r_l \hat r_m \hat r_n]\;,
\eea
with
\bea
U_1(x)&=& x^4 + 2x^3 + 3x^2 + 3x +3\;,\\
U_2(x)&=& -x^4 - 2x^3 - x^2 + 3x +3\;,\\
U_3(x)&=& x^4 + 2x^3 - 3x^2 - 15x -15\;,\\
U_4(x)&=& -x^4 - 4x^3 - 9x^2 - 15x -15\;,\\
U_5(x)&=& x^4 + 10x^3 + 45x^2 + 105x +105\;.
\eea
For a linear arrangement with $r_A=2r_B=2r_C=2r$, the far-zone three-body interaction potential Eq. (\ref{Eabc far}) becomes
\beq\label{Eabc far1}
\Delta E_{ABC}\approx-\frac{0.1}{\pi^4 r^{16}} \alpha^{A}(0) \alpha^{B}(0) \alpha^{C}(0)\;.
\eeq
Obviously, the three-body quantum gravitational quadrupole interaction shows an $r^{-16}$ dependence in the far regime and is attractive since the sign of the potential Eq. (\ref{Eabc far1}) is negative. For the triangle configurations, we also consider two examples with special internal angles to show the geometrical dependence of the attractive or repulsive properties of the far-zone three-body interaction.
For an equilateral triangle in which $r_A=r_B=r_C=r$, the far-zone interaction potential Eq. (\ref{Eabc far}) becomes
\beq\label{Eabc far2}
\Delta E_{ABC}\approx-\frac{7.5}{\pi^4 r^{16}} \alpha^{A}(0) \alpha^{B}(0) \alpha^{C}(0)\;,
\eeq
whereas for a triangle in which $r_A=r$ with $\theta_A=\frac{2}{9}\pi$, $\theta_B=\frac{\pi}{9}$, $\theta_C=\frac{2\pi}{3}$, the far-zone potential Eq. (\ref{Eabc far}) is
\beq\label{Eabc far3}
\Delta E_{ABC}\approx\frac{2.4}{\pi^4 r^{16}} \alpha^{A}(0) \alpha^{B}(0) \alpha^{C}(0)\;.
\eeq
where the relation $\frac{r_A}{\sin{\theta_A}}=\frac{r_B}{\sin{\theta_B}}=\frac{r_C}{\sin{\theta_C}}$ has been utilized.
Obviously, the distance dependence of the far-zone three-body quantum gravitational interaction in the triangle configurations is still $r^{-16}$. Moreover, the sign of Eq. (\ref{Eabc far2}) is negative so that the corresponding force is attractive, while that of Eq. (\ref{Eabc far3}) is positive and therefore the force is repulsive. Thus, in the far regime, the three-body quantum gravitational quadrupole interaction can also be attractive or repulsive depending on the geometrical configuration of the three objects, which is similar to the case in the near regime.

\section{Discussion}
\label{sec_disc}
%\setcounter{equation}{0}
%%%%%%%%%%%%%%%%%%%%%%%%%%%%%%%%%%%%%%%%%%%%%%%%%%
In this paper, we study the nonadditive three-body quantum gravitational quadrupole interaction between three ground-state nonpointlike objects coupled with the fluctuating gravitational fields in vacuum, based on the perturbation theory in the framework of linearized quantum gravity. Our result shows that the three-body quantum gravitational interaction displays a  distance behavior of  $r_A^{-5}r_B^{-5}r_C^{-5}$ in the near regime where the interobject distances are smaller than the characteristic transition wavelength of the objects, and  $r_A^{-5}r_B^{-5}r_C^{-5}(r_A+r_B+r_C)^{-1}$ in the far regime where the distances are larger than the characteristic transition wavelength. Here $r_\xi $ denotes the distance between any two objects except object $\xi\;(\xi=A,B,C)$.
We also find that, both in the near and far regimes, such a three-body interaction is attractive for a linear arrangement and can be attractive or repulsive depending on the internal angles for a triangle configuration. That is, the attractive or repulsive properties of the nonadditive three-body quantum gravitational quadrupole interaction can in principle be manipulated by changing the geometrical configuration of the three objects, in contrast to the additive two-body case which is always attractive.

\begin{acknowledgments}

We would like to thank Jiawei Hu for helpful discussions.
This work was supported in part by the NSFC under Grants No. 11690034 and No. 12075084.
\end{acknowledgments}

\end{document}